\documentclass{kluwer} 
\usepackage{epsfig} 
\begin{document} 
\begin{article} 
\begin{opening} 
 
\title{H$_\alpha$ Luminosity and Star Formation of  
Galaxies \\ in HCGs} 
\author{Paola \surname{Severgnini}\email{paola@ifctr.mi.cnr.it, paolas@arcetri.astro.it}} 
\institute{IFC-CNR 'G. Occhialini', via Bassini 15, 20133 Milano, Italy \hskip 4truecmUniv. Studi di Firenze, Dip. Astronomico di Arcetri, Firenze, Italy}
\author{Paolo \surname{Saracco}\email{saracco@merate.mi.astro.it}} 
\institute{Osservatorio Astronomico di Brera, via E. Bianchi 46, 22055 Merate, Italy} 
 
\runningtitle{Star Formation in HCGs} 
\runningauthor{P. Severgnini \& P. Saracco} 
 
\begin{ao} 
Paola Severgnini\\ 
IFC 'G. Occhialini', via Bassini 15\\ 
20133 Milano, Italy\\ 
e-mail: paola@ifctr.mi.cnr.it 
\end{ao} 
 
\begin{abstract} 
 
We have obtained  $H_{\alpha}$ fluxes and luminosities for a sample of 95  
accordant galaxies from observations of 31 Hickson Compact Groups (HCGs) 
in the North hemisphere. 
This sample is  the largest $H_{\alpha}$ selected catalogue  
of galaxies having  $H_{\alpha}$ calibrated fluxes so far. 
The results obtained  from a preliminary analysis  of a subsample of 66 galaxies show that the $H_{\alpha}$ luminosity of galaxies
 is correlated with velocity dispersion and  compactness of groups. 
Such correlations   would point toward a scenario in which $H_{\alpha}$ 
brightest galaxies reside in compact groups having higher probability 
of galaxy interaction i.e. lower values of velocity dispersion. 
Moreover such relations seem to depend on the environment in which 
HCGs themselves are embedded. 
 
\end{abstract} 
\end{opening} 
 
\section{Introduction} 
Hickson Compact Groups (hereafter HCGs; Hickson 1982) are small systems of  
several galaxies (four or more) in an apparent close proximity in the sky. 
Their dynamical state has been the subject of a controversial debate. 
Comparison of the observational data with model calculations led Mamon 
(1986, 1987) to conclude that less than a half of HCGs could be 
considered bounded dense systems. 
However accumulated statistical evidences (Hickson \& Rood 1988;  
Hickson 1992) together with observational evidences, also coming from X-ray 
observations (Saracco \& Ciliegi 1995; Ponman et al. 1996; Pildis et al. 1995),
 clearly favor the view that the majority of HCGs are physical systems 
and not chance projections or transient systems. 
 
One of the expected consequences of the gravitational interactions 
between galaxies is the enhancement of the star formation rate (SFR) 
in the interacting systems (Joseph \& Wright 1985; Bushouse 1987; 
Laurikainen \& Moles 1989). 
The photometric and spectroscopic studies carried out so far  
(Rubin et al. 1991 \& Mendes de Oliveira et al. 1997; Moles et al.  
1994 \& Mendes de Oliveira et al. 1994; Vilchez \& Iglesias Paramo 1998; 
Plana et al. 1998, Iglesias Paramo \& Vilchez 1999),  
aiming at establishing the fraction of interacting galaxies in HCGs,  
have often given contradictory results. 
Actually a possibility exists that only a fraction of the  HCGs are bound  systems and that bound HCGs evolve in different ways. 
Different evolutions could be due to both the different dynamical  
properties of HCGs and to  their different 'birthplace', i.e.  
the environment in which they are embedded. 
 
Powerful signs  of  star formation activity are the  
ionization lines emitted by the heated gas surrounding the regions of star  formation. 
Unlike the color indexes in the $U,B,V$ filters, that give indications  
about the past star formation  ($> 10^8$ years), 
the H$_\alpha$ emission line at 6563 {\AA} can be used as a quantitative  
and spatial tracer of the rate of massive ($\geq$ 10 M$_\odot$) and  
therefore recent ($\leq 10^7$ years) star formation (Kennicutt 1983;  
Ryder \& Dopita 1994). 
Therefore, knowing the  H$_\alpha$ emission of the HCG galaxies, it is   
possible in principle to carry out important knowledges about the  present  
merger and interaction events in these systems. 
Up to now the only H$_\alpha$ images regarding HCG galaxies have been 
collected by Rubin et al. (1991) and more recently by Vilchez \& Iglesias 
Paramo (1998).  
They carried out H$_\alpha$ emission-line images respectively for 14 and  
16 HCGs. While Vilchez \& Iglesias Paramo estimate the H$_\alpha$ flux for each of the 63 galaxies of their sample (Iglesias Paramo \& Vilchez 1999), Rubin et al. do not use flux calibrated and they take into account a sample constituted by disk galaxies only.
We have recently obtained  $H_{\alpha}$ fluxes and luminosities for  
a sample of 95  
galaxies from calibrated observations of 31 HCGs (Severgnini et al. 1999). 
Here we present the preliminary results of the analysis performed on  
a subsample of 66 galaxies. 
 
\section{Observations} 
 
Observations have been carried out at the 2.1 meter telescope (design  
Ritchey-Chretien) at the National Observatory of Mexico in S.  
Pedro Martir during three different observing runs (November 1995,  
April 1996 and February 1997).  
The telescope was equipped with a Tektronix CCD of 1024x1024 pixels,  
each 24$\mu$m x 24$\mu$m. The telescope scale (13 arcsec/mm) and the  
pixel dimensions provide a pixel size of 0.3 arcsec/pix with a resulting  
field of view of 5.12$^\prime\times$5.12$^\prime$. 
During these three runs we observed 31 HCGs in H$_\alpha$ filters. 
The remaining 61 HCGs were not in ours sample because the adequate H$_\alpha$  
interferometric filters were not available.  
This is the only criterion used to select the observed groups.\\ 
 
In order to calibrate our data, we have observed some spectrophotometric  
stars, equally spaced in time during each night, taken  
 from the list of Massey $\&$ Strobel (1988).  
The standards were observed in all the H$_\alpha$ narrow-band   
filters used to observe HCGs. 
 
During the observations the seeing was in the range of 2 to 2.6 arcsec  
and the photometry was within 0.05 mag in all but one (worse) night. 
The mean limiting flux of the observations, at one sigma from the  
background and within the mean seeing disk (2.3 arcsec), is  
9.22$\cdot10^{-17}$ erg cm$^{-2}$ s$^{-1}$. 
We estimated the $H_{\alpha}$ flux and  
luminosity for 66 galaxies, 12 out of which are upper limits. 
We adopt H$_0$=100 km/(s Mpc) and q$_0$=0.5.\\ 
In Figure 1 the distribution of $H_{\alpha}$ luminosity ($L_{H_\alpha}$) of 
the 54 detected galaxies is shown. $L_{H_\alpha}$ for each galaxy has been 
derived from using a $H_\alpha$ isophotal flux computed within the region defined 
by a detection threshold of one sigma above the background.
A detailed description of the data reduction, the photometric calibration, the flux estimate and the luminosity derivation is given in Severgnini et al. (1999). 
\begin{figure} 
\centerline{\epsfig{file=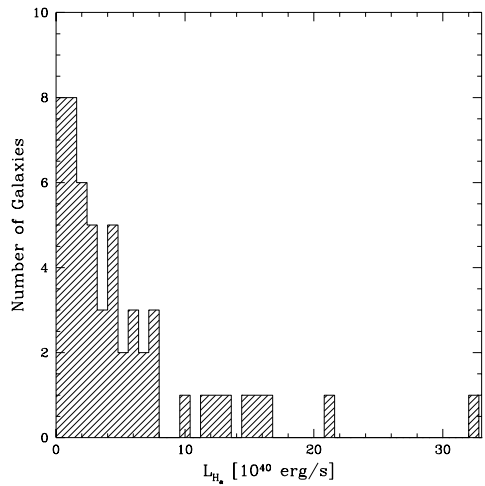, width=16pc}} 
\caption{Distribution of $H_{\alpha}$ isophotal luminosity, corrected for Galactic 
Extinction (see Severgnini et al. 1999),  of the 54 detected galaxies.} 
\end{figure}

\begin{figure} 
\centerline{\epsfig{file=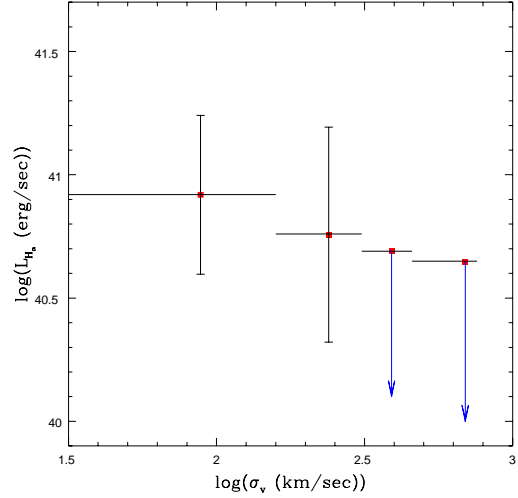, width=16pc}} 
\caption{$H_{\alpha}$ luminosity of the 66 galaxies of our sample vs velocity  
dispersion of the groups. There is a clear correlation (the probability 
of the two not being correlated is P=0.001) which suggests 
an increasing of $H_{\alpha}$ luminosity of galaxies with decreasing 
of velocity dispersion of groups.} 
\end{figure} 
 
\begin{figure} 
\centerline{\epsfig{file=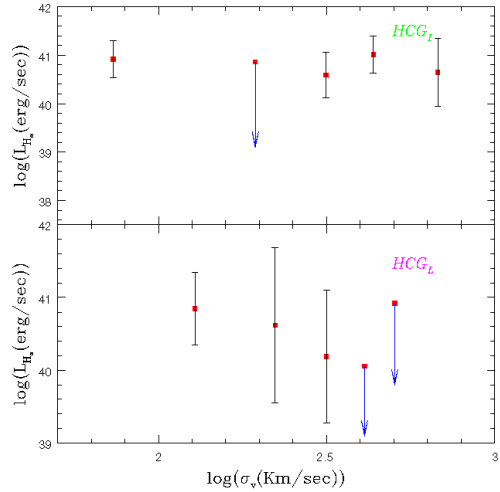,width=16pc}} 
\caption{$H_{\alpha}$ luminosity of galaxies vs velocity  
dispersion of the isolated HCGs ({\it upper panel}) and of the HCGs embedded in loose groups ({\it lower panel}). The velocity dispersion of HCG$_L$  is correlated to H$_\alpha$ luminosity of their galaxies  
(the probability of the   
two not being correlated is P=0.03), 
while the same correlation is absent for HCG$_I$ (P=0.60). 
In both the Figures 2 and 3, galaxies are binned in dynamical properties. 
The dots represent the mean luminosity and the error bars are 
the standard deviations of the values. 
The down arrows refer to upper limits inside  
of the bin. In Figure 2 the width of the bins is also  
shown.} 
\end{figure}

\section{Results}  

We have investigated about the dependency of  the
H$_\alpha$ luminosity  of galaxies on the dynamical properties 
of  HCGs  and on the  environment in which groups are embedded.
The statistical analysis we have performed has made use of the
survival  analysis (Isobe, Feigelson \& Nelson 1986) since
some of the observed galaxies have not been detected in our observations.

In an undisrupted galaxy it is expected that actual SFR ($<$10$^{6-7}$ years) 
is  correlated to the quantity of young stellar population (O,B stars). 
To test if this correlation is present also for HCG galaxies we searched 
for a  correlation between the $H_{\alpha}$ luminosity of galaxies and 
their $B$ absolute magnitude (M$_B$ ({\em "Atlas of Compact Groups of  
Galaxies"}, Hickson 1993)), good tracer of young stars. 
We found a significant correlation between these two  quantities 
in the sense that:  
\begin{description} 
\item $\bullet$  galaxy having higher (lower) $H_{\alpha}$ luminosity have  
higher (lower) $B$ luminosity.  
\end{description}

We  have then searched for correlations between the dynamical  
parameters of HCGs (velocity dispersion, mass density, surface 
brightness and crossing time, as from Hickson et al. 1992) and 
the $H_{\alpha}$ luminosity of the member galaxies.  
Our analysis shows the presence of correlations between each of 
these parameters  and the  
$H_{\alpha}$ luminosity of the galaxies. 
In particular we found that the   
$H_{\alpha}$ emission is correlated to crossing time and  it  
is anti-correlated to velocity dispersion, mass density and surface  
brightness in the sense that:
\begin{description} 
\item $\bullet$ HCGs having higher (lower) densities have  higher (lower)  
velocity dispersions and the galaxies inside them have lower (higher)  
$H_{\alpha}$ luminosities. 
\end{description} 
In Figure 2  the relation  between $H_{\alpha}$ luminosity   
of galaxies and the velocity dispersion of the groups is shown  
(the probability of the two not  being correlated  
is 0.001). 
Such correlations would seem to point toward 
a dependence of the   $H_{\alpha}$ luminosity of galaxies,
and hence of their SFR, on the dynamical properties of HCGs.


To test if different environments, where HCGs are actually 
found, affect the H$_\alpha$ of galaxies inside them we have divided  
our sample  in two subsamples, following the 
Rood and Struble's (1994) classification:  
the first composed by the isolate compact groups (HCG$_I$) and the second one  
consisting of those compact groups embedded in  loose groups (HCG$_L$). 
Through a statistical comparison we found that there aren't significant    
differences between  $H_{\alpha}$ luminosity distributions of galaxies inside  
HCG$_I$ and HCG$_L$. The same result is found if we compare  
the dynamical parameters of isolate HCGs to those of HCGs in loose groups. 
Nevertheless, from our data we found that the velocity dispersion (Figure 3) and  
crossing time of HCG$_L$  are correlated to the $H_{\alpha}$ luminosity of their  
galaxies and that these correlations are absent for HCG$_I$. 
We thus can assert that inside HCG$_L$ there are correlations between two  
dynamical parameters and $H_{\alpha}$ emission that are not present  
inside HCG$_I$. 
These results suggest that, although the surrounding environment of HCGs does not  
directly influence the SFR of galaxies and the dynamical properties of groups, it  
modifies the relation between  dynamical properties of groups and SFR of their  
galaxies.\\ 
Moreover we have tested that the correlation found between $H_{\alpha}$  
and $B$ luminosity is present only for galaxies inside HCG$_L$ (Figure 4). This  
result suggests that the different HCG environments affect also 
the ratio between the population of young stars  and actual star formation of  
galaxies.\\ 
Thus we  assert that the surrounding environment of HCGs affects the evolution of  
the HCG galaxies. 
 
\begin{figure} 
\centerline{\epsfig{file=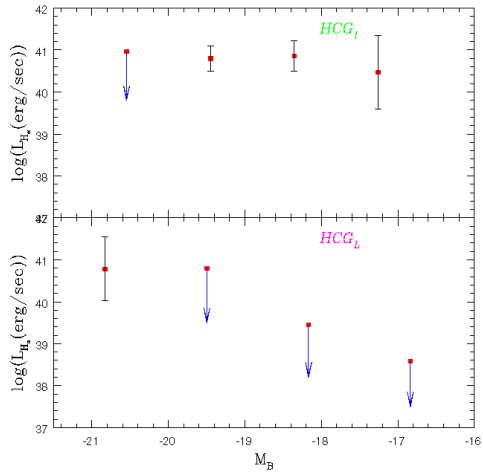, width=16pc}} 
\caption{ {\it Upper panel:} The H$_\alpha$ luminosity of galaxies in HCG$_I$ vs their own absolute magnitude in the $B$ filter ($M_B$).
{\it Lower panel:} The H$_\alpha$ luminosity of galaxies in HCG$_L$ vs their own absolute magnitude in the $B$ filter ($M_B$).
$M_B$ of galaxies into HCG$_L$ is correlated to their own H$_\alpha$ luminosity (the probability of the two not being correlated is P=0.007) while such correlation is absent for HCG$_I$ (P=0.17).
In both the figures, galaxies are binned in  $M_B$ values.}
\end{figure} 
Finally, we also investigated if the morphology  of galaxies is influenced by  
different environments and if a particular HCG surrounding environment can favor  
the formation and evolution of a particular morphology. 
We divide all the galaxies of Hickson's sample in Ellipticals $+$  
Lenticulars (E/S0) and Spirals $+$ Irregulars (S/I) finding  the same fraction of  
E/SO and S/I inside HCG$_I$ and  HCG$_L$ respectively. 
Thus we conclude  that the  environment in which HCGs are embedded  does not  
influence the morphology of their member galaxies.

\section{Discussion and Conclusions} 
We have studied H$_\alpha$ calibrated fluxes for a sample of 66 galaxies 
in 31 HCGs.
12 galaxies of our sample have not been detected in our images and 
this implies that they have  a  H$_\alpha$ emission lower than  
$f_{lim}=$9.22$\cdot 10^{-17}$ erg s$^{-1}$ cm$^{-2}$, being $f_{lim}$  
the 1$\sigma$ limiting flux integrated within one seeing disk reached in  
our observations. 
 
From the analysis presented above, we could conclude that the $H_{\alpha}$ 
luminosity 
of galaxies is affected by the dynamical properties of compact groups.
The correlations found would point towards a merger scenario in which  
$H_{\alpha}$ brightest galaxies are inside groups with higher probability  
of interaction i.e. lower values of velocity dispersion and mass density. 
It is worth noting however that $H_{\alpha}$ luminosity
could be matched by  mass of  galaxies, that is the
SFR could be higher (lower) because of the higher (lower)
quantity of  the available gas 
and not because of a real higher {\em efficiency} or {\em fraction}.
In principle such degeneracy could be avoided  normalizing the 
$H_{\alpha}$ luminosity by a mass tracer of galaxies which is very 
well represented by the luminosity in the near IR band (H or K band), 
as shown by Gavazzi et al. (1996).
Nevertheless some authors (e.g. Iglesias Paramo et al. 1999) make use of the
absolute B magnitude  under the implicit assumption that $M\propto L_B$.
On the other hand, since a correlation between $H_{\alpha}$ and B luminosities
is expected under robust assumptions (and we found it) the use of $L_B$ 
to remove the degeneracy could lead to misleading results.
Thus, even if we find correlations between $H_{\alpha}$ luminosity
and dynamical parameters of the groups, further investigations are required
in order to establish if  the dynamics of HCGs affects  the evolution of their member galaxies.

Moreover the relations we find  seem to depend  
on the environment in which HCGs are embedded, as confirmed by the  
correlations found only in the case of HCG$_{L}$ sample.  
Since the $H_{\alpha}$ luminosity  
is a good star formation tracer, this result  suggests that also 
the environments surrounding HCGs could influence the evolution  
of galaxies into HCGs. 
 
The SFR for objects can be directly inferred  by H$_\alpha$ luminosity 
using the result of Kennicut (1983), which 
relates the SFR to $H_{\alpha}$ luminosity through the relation 
\begin{equation} 
SFR(total)={{L(H_\alpha)}\over{1.12\cdot10^{41} {\rm erg~~s^{-1}}}} M_\odot yr^{-1} 
\end{equation} 
where a Salpeter initial mass function with an upper mass cutoff of 100 
$M_\odot$ have been assumed. 
From the luminosities we infer a star 
formation rate for our detected galaxies in the range 0.01-2.88 $M_\odot$ yr$^{-1}$ in absence of internal extinction.

\end{article} 
 
\end{document}